# The Gompertz Growth of COVID-19 Outbreaks is Caused by Super-Spreaders

Francesco Zonta[2], Andrea Scaiewicz[1], and Michael Levitt[1*]

[1] **Department of Structural Biology, Stanford University School of Medicine, Stanford, CA 94305, USA.**
[2] **Shanghai Institute for Advanced Immunochemical Studies, ShanghaiTech University, Shanghai, China**

*Corresponding Author: michael.levitt@stanford.edu

**ABSTRACT**

In individual SARS-CoV-2 outbreaks, the count of confirmed cases and deaths follow a Gompertz growth function for locations of very different sizes. This lack of dependence on region size leads us to hypothesize that virus spread depends on universal properties of the network of social interactions. We test this hypothesis by simulating the propagation of a virus on networks of different topologies. Our main finding is that Gompertz growth observed for early outbreaks occurs only for a scale-free network, in which nodes with many more neighbors than average are common. These nodes that have very many neighbors are infected early in the outbreak and then spread the infection very rapidly. When these nodes are no longer infectious, the remaining nodes that have most neighbors take over and continue to spread the infection. In this way, the rate of spread is fastest at the very start and slows down immediately. Geometrically it is seen that the "surface" of the epidemic, the number of susceptible nodes in contact with the infected nodes, starts to rapidly decrease very early in the epidemic and as soon as the larger nodes have been infected. In our simulation, the speed and impact of an outbreak depend on three parameters: the average number of contacts each node makes, the probability of being infected by a neighbor, and the probability of recovery. Intelligent interventions to reduce the impact of future outbreaks need to focus on these critical parameters in order to minimize economic and social collateral damage.

**DEFINITIONS**

| | |
|---|---|
| $M$ | total number of nodes in the network |
| $X(t)$ | total number infected at time $t$ |
| $N = \lim_{t \to \infty} X(t)$ | plateau value for the number of infected |
| $P_I$ | probability of infection (per unit of time) |
| $P_R$ | probability of recovery (per unit of time) |
| $R(t)$ | number of recovered at time $t$ |
| $I(t) = X(t) - R(t)$ | number of people that can infect other people at time $t$ |
| $\Delta(t) = X(t) - X(t-1)$ | new infections at time $t$ |
| $f(t) = \dfrac{X(t)}{X(t-1)}$ | fractional change |
| $H(t) = \ln[f(t)] = \ln\left[\dfrac{X(t)}{X(t-1)}\right]$ | |
| $J(t) = \ln[H(t)] = \ln\left[\ln\left[\dfrac{X(t)}{X(t-1)}\right]\right]$ | |
| $\Sigma(t)$ | epidemic surface (total number of contacts between infected and susceptible nodes - multiple infected nodes may be in contact with the same susceptible node) |

## INTRODUCTION

A careful analysis of SARS-CoV-2 individual clearly resolved outbreaks in 113 world locations shows that in single isolated outbreaks, the generally spreads according to the Gompertz growth function (1-4). This simple function is characterized by three parameters: $N$, $U$ and $T$, so that total counts, $X(t)$ at time $t$ is

$$X(t) = Ne^{e^{-\frac{(t-T)}{U}}} \qquad (1).$$

where $N$ is the total size of the outbreak, $U$ is a time constant that determines how fast the growth and the slowdown occur, and $T$ is the day the number of daily counts peaks.

In most locations, the number of deaths associated with the virus follows the same trajectory as the total number of cases, with approximately the same $U$ value.

Analogous behavior can be found in other respiratory disease epidemics, for example the 2003 SARS outbreak (5), suggesting that respiratory viruses spread following simple and universal rules that are largely independent of genetic or social differences in the various countries.

In this paper, we propose a simple model of virus diffusion on networks (6, 7), which represent normal human social interactions. We show that it is possible to reproduce actual data of isolated outbreaks, provided that the network is scale-free. The model also allows us to derive a causal relationship between the total number of infections and some "microscopic" parameters that are properties of the virus and its interaction with society. Social intervention can change these parameters, and understanding their relative effect on the spread of the virus can help determine the most cost-effective strategy to contain virus spread in future outbreaks.

## RESULTS

**A simple model of virus diffusion on a scale-free network reproduces Gompertz dynamics found in observed data.**

We considered a simple infection model on a network and studied its evolution. In our model, individuals are represented as nodes connected by links representing the possibility of interaction (and thereby virus spreading). If node $i$ interacts with node $j$, node $j$ also interacts with node $i$ (undirected network). The number of links $k$ that are connected to a particular node is called the *degree* of the node. We assume links do not change with time so that the network is static. One can interpret links as connections between people whose interaction is sufficient to allow infection.

We follow the dynamics of the spreading of the virus using a discrete version of the **SIR** model (6, 8) in which each node can be in one of three states: State **S** defines a *Susceptible* node, which can become *Infected* (State **I**) with a given probability $P_I$, if and only if it is connected to another infected node. Infected nodes can become *Recovered* (State **R**) with a probability $P_R$, and they cannot be re-infected. These 'immune' nodes represent individuals who can no longer infect or be infected, either because they have successfully recovered from the disease, have been isolated from the network (for example, placed in hospitals or quarantine), or are no longer alive. The infection propagates on the network according to the following algorithm. At time $t=0$, a small number ($m_0$) of nodes are randomly infected. At each subsequent time step, an infected node infects a linked node with a probability $P_I$ and recovers with a probability $P_R$. The process is iterated until reaching an equilibrium state in which there are no more infected nodes (**Figure 1A**).

We observe that the shape of the virus spread in the network depends on the network topology: if the network is *scale-free (9-11)*, i.e., the distribution of the degrees of the nodes follows a power law $P(k) = k^{-\gamma}$ with $2 < \gamma < 3$ (**Figure 1B**), the propagation of the virus follows a Gompertz law (**Figure 1C**) and with appropriate choice of parameters can reproduce remarkably well actual observed data (**Figure 1 panels D and E** show a comparison with cases in Italy cases, after rescaling). Following Gompertz and fitting real-world data is a characteristic of the scale-free network: other topologies (small world or random networks) fail to produce the same behavior (**Fig. 5B & Supplementary Figure S1**).

Due to the stochastic nature of our model, we still observe situations where the spread of the virus does not follow a Gompertz growth function: these situations always correspond to simulations with microscopic parameters $P_R$ and $P_I$ in which the virus never infects more than a few nodes (up to 50) before disappearing. Such cases become more probable as the ratio between $P_R$ and $P_I$ increases but occur in less than 30% simulations even when $P_R/P_I = 5$. In other words, when the virus actually spreads through the network, it always does so by following the Gompertz growth function.

**In a scale-free network, outbreaks are entirely determined by microscopic parameters and network connectivity.**

Having shown that when outbreaks are significant, they always follow the Gompertz growth function, we wanted to understand how the parameters $N$ and $U$ in Eq. 1 are linked to the microscopic rate of infection and recovery and possibly to other network properties. We do not consider the third parameter ($T$) because it merely causes translation of the Gompertz curve along the time axis.

To study the effect of microscopic parameters, we run many simulations with different values for $P_I$ and $P_R$ on the same network, varying the initial seeding of the infection. Similarly, we tested how the results depend on the parameters that define the network (namely, the total number of nodes M and the average connectivity $\bar{k}$). In order to have a rough comparison with actual data, simulations were run with $P_I$ and $P_R$ varying in the range 0.01 – 0.2 (see Methods).

Despite our model being stochastic, we find that the percentage of the total nodes ($m=N/M$ where $M$ is the size of the network) and the time-constant ($U$) are well-defined functions of the two parameters $P_I$ and $P_R$ as well as of the average connectivity of the network $\bar{k}$. This allows these quantities to be predicted quite well without simulation (to within small statistical fluctuation). In particular, $U$ is inversely proportional to the product of $\bar{k}$ and $P_I$, ($U = 1/\alpha \bar{k} P_I$) while $m$ is a decreasing function of $x = U \cdot P_R = P_R / \alpha \bar{k} P_I$ (**Figure 2**):

$$U = 1/\alpha \bar{k} P_I \qquad (2)$$

$$m = m(x) = \sum_{i=1}^{\infty} a_i e^{c_i x^i} \qquad (3)$$

$m(x)$ is a monotonically decreasing function of $x$; it does not have a simple, functional form but can be well approximated by the first four terms of the series in Equation 3 (see fit in **Figure 2**). Constants $\alpha$, $a_k$, and $c_k$ in Equations 2 and 3 are fitting parameters and do not depend on the size of the network. As shown in **Figure 2 panel D and E,** when $P_R/P_I < 0.5$, $m$ is close to 1 and the network is almost entirely infected. Taking into account Equation (2) and (3), we can rewrite Eq (1) as:

$$X(t) = M n\left(P_R / \alpha \bar{k} P_I\right) \cdot e^{e^{-\alpha \bar{k} P_I (t-T)}} \qquad (4)$$

Equation 4 shows that it is possible to describe viral spread on a scale-free network from microscopic parameters that could be available *a priori* from intrinsic, known characteristics of both the virus and the network.

**Infection of major hubs**

We further analyzed the role of major hubs, the nodes with many neighbors, $k$, the super-spreaders) in spreading the virus. **Figure 3 panel A** gives the time, $t_0$, at which each node of the network is infected as a function of $\ln[1/k]$, averaged over 100 simulations run with $P_I = 0.01$ and $P_R = 0.01$ using different random seedings of the infection. Major hubs with high k values are consistently infected at shorter $t_0$ than smaller hubs. The smaller hubs also show much larger differences in $t_0$ than the larger hubs. Similar trends can be observed for other choices of parameters (**Figure 3 Panel B** and **Supplementary Figure 2**). The time to be infected is largely independent of $P_R$, unless the ratio $P_R/P_I$ is close to 1.

To understand whether the infections of major hubs actually cause faster spreading of the virus, we repeat the simulation with the additional condition that the 20 largest hubs (corresponding to nodes with $k>100$) cannot be infected. **Figure 4** shows histograms for the fractional plateau value $m$ for those simulations where the virus spreads widely in the network. We see that while the 'immunity' of the largest nodes make negligible difference when $P_R/P_I=1$, it becomes important when $P_R/P_I=5$. In addition to the average value of $m$ being smaller when big hubs are immune, the number of cases in which spread does not occur increases from 29% to 48% for $P_R/P_I=5$, while it is almost unaltered – 1% to 2% – when $P_R/P_I=1$ (**Figure 4 inset**). These numbers are for simulations in which 4 different random nodes are infected at $t=0$; we expect more significant differences if the initial seeding number is smaller. Overall, we conclude that while infection of the nodes larger than 100 is not essential for the propagation of the epidemic, they are important drivers of virus spread by influencing the probability that an outbreak actually spreads and to what fraction of the network.

**In a scale-free network, the "surface" of the epidemic rapidly decreases after the largest nodes have been infected**

The Gompertz function was initially envisioned to describe growth in an environment with limited resources (1). Its most important feature is that the Gompertz growth function starts with maximum velocity and then slows down immediately. Why does this happen in a scale-free network and not in other networks?

To answer this question and understand why this is a special property of the scale-free network, we track the time course of several observables and determine how they are correlated. Results are reported in **Figure 5, panels A** for three different replicas of infection of a scale-free network and **panels B** for one example of infection in a random network.

The top row panels show functions $X(t)$ (red line, total number of nodes that have been infected), $I(t)$ (gray line, number of infected nodes that did not recover yet), and $\Delta(t)$ (orange line, daily new infected nodes, shown on a different scale for clarity). We can immediately observe that $I(t)$ and $\Delta(t)$ lose correlation after some time which happens much faster for a scale-free topology.

$\Delta(t)$ is proportional to $\Sigma(t)$, the number of contacts that infectious nodes make with non-infected nodes (the "surface" of the epidemic, see Methods section). The fact that $\Delta(t)$ decays faster than $I(t)$ means that nodes are infected later on, have less access to the pool of susceptible nodes. For a random network, this decay happens halfway through the infection, when the susceptible nodes are comparable or less than the infected nodes, as one would expect from a logistic growth; but happens much earlier in a scale-free network.

This can be understood by observing that in a scale-free network, bigger nodes are exhausted very early in the epidemic (**Figure 4 and Figure 5 central rows**), and thus the number of contacts that can be established with uninfected nodes immediately decays.

By contrast, in a random network, the biggest nodes are not very different from the others. While on average, they are also infected faster and more consistently, they do not appear to have any particular role in spreading of the epidemic.

Infection of largest nodes in a scale-free network coincides with the beginning of the Gompertz growth of the epidemic, as is evident from the linear behavior of $J(t)$ (**bottom rows in Figure 5)**. By contrast, in a random network, such nodes do not seem to have any special role.

**Multiple outbreaks cannot be observed in a single scale-free network, but emerge in multiple networks connected by a few links.**

In contrast with reality, we never observe multiple outbreaks in single scale-free or random networks: in all the simulated trajectories, the virus exhausts itself following the Gompertz growth function (or other growth functions for the random network) even if the final fraction of the network that has been infected is small. Indeed, as shown in **Figure 4 and Figure 5**, larger nodes, which fuel the Gompertz outbreak, are always infected early, and after they recover, there is no possibility of creating a new outbreak.

However, we can observe a second Gompertz outbreak if we connect two different scale-free network with a small number of edges (**Figure 6**). The outburst of the second outbreak can happen almost immediately or well after the first outbreak has passed its peak, and the overall dynamics are reminiscent of real-world data (to be released soon in another paper).

**DISCUSSION**

The virus spread model we analyzed in this work correctly reproduces the main characteristics of COVID-19 in the real world. At the beginning of the infection, the virus propagates in the network infecting a small number of other nodes. This phase is likely to go undetected in the real world, as the number of infected individuals is small and COVID-19 can be asymptomatic. Depending on the characteristics of the virus, the infection can either disappear during this first phase or last enough to reach one of the major nodes. In a scale-free network, such nodes are much better connected than the average node, and thus they are typically infected very early. When one of such nodes is infected, we observe a rapid increase in the number of infected nodes, resulting in a major outbreak which will hide any smaller or secondary outbreak, i.e. the network is synchronized once major nodes are involved in the infection. The largest nodes propagate the infection towards other nodes that are smaller but still better connected than the rest of the network, and the process iterates through smaller and smaller nodes. At a certain point, depending on how fast nodes can recover, the infection self-limit because the number of nodes connected to each infected node becomes smaller and smaller.

The existence of super-spreaders and the organization of the network in a scale-free topology is needed to correctly reproduce the Gompertz growth of the virus. While this does not necessarily imply that all human interaction in society is organized according this particular topology, as scale free networks are indeed expected to be rare (12, 13): it indicates how society is organized (or perhaps how the virus sees society) on some intermediate scale, for example at a city level. This also gives indirect evidence that the virus is airborne (14) since no other transmission routes can explain why some individuals could infect hundreds or thousands of people in a short amount of time (15).

**Effect of intervention in the virus propagation.**

**Equation 3** shows that the time constant $U$, which characterizes the speed at which the infection propagates in the network, depends only on two factors. One is the infection probability ($P_I$), which is a characteristic of the virus. The other is the average degree of connectivity of the social network of human interaction in a city, region, or country ($\bar{k}$), which summarizes complex interaction patterns between people.

$P_I$ can be different for different virus strains and can perhaps be modulated by the ambient temperature (the season) or by using masks, visors, gloves, and other protective measures. Similarly, the average connectivity can be modified by spontaneous human behavior or by government intervention.

The fraction of the network infected by the virus ($m$) depends on $U$ and the recovery rate probability ($P_R$). $P_R$ represents the rate at which infected nodes recover (or die) or are isolated from the network. Its value does not depend only on the virus characteristics but can be influenced by social interventions such as isolating infected people.

Increasing $P_R$ by a certain factor is equivalent to decreasing $P_I$ or $\bar{k}$ by the same factor. However, the methods to achieve either case are drastically different in the real-world: increasing $P_R$ can be done a posteriori by tracking and isolating infected people while decreasing $\bar{k}$ would require a priori intervention such as lockdowns or school closure. The feasibility and the economic cost of the two approaches are different and depend on the country's internal policies.

Another way to affect the final number of infected people is by capping the extent of the network that the virus has compromised. In our model, this is represented by the size $M$ of the network, and is a fixed number. Nevertheless, in reality, it may be possible to isolate the virus before it expands and infects a new "region" of the network.

**CONCLUSIONS**

Despite being minimal, the model proposed in this paper reproduces most of the characteristics of COVID-19 outbreaks. It successfully reproduces the observed Gompertz growth behavior, explains why outbreaks present universal characteristics, and most importantly, provides a theoretical framework to interpret differences between various countries as differences in the social structure or how society responds to the virus. It is undoubtedly possible to improve the realism of the model in various ways, for example, allowing for a dynamic contact structure. We believe that such changes may improve the description of what happens between two outbreaks but will not affect the modeling of single outbreaks. Indeed, the initial phase, in which the virus lingers within the network before finding one of the larger nodes, appears to be a stochastic process that is likely to be largely independent of the network properties; while the rapid growth of the Gompertz phase happens so fast that the underlying network would not have time to change.

# METHODS

**Simulations and computer code.**

The original code for generating the network and the stochastic model of virus diffusion is written in C. Results have also been double-checked and reproduced using a Python script and the Networkx package (16).

**Choice of parameters for simulations.**

It is possible to roughly estimate realistic values for $P_R$, as it represents the probability that a person recovers from the virus in a given time interval.

The average time that a node remains infected can be calculated analytically from $P_R$ as follows.

$P(n)$ the probability of recovering by day $n$, can be written:

$$P(1) = P_R \cdot (1 - P_R)^0$$
$$P(2) = P_R \cdot (1 - P_R)^1$$
$$\ldots$$
$$P(n) = P_R \cdot (1 - P_R)^{n-1}$$

Now the average time can be computed as:

$$\langle n \rangle = \sum_{d=1}^{\infty} n P_R \cdot (1 - P_R)^{n-1} = \frac{1}{P_R}$$

We know from known cases of COVID-19 that the recovery time is about one week but can take up to 2 months. For this reason, we choose $P_R$ values in the range 0.01 to 0.2 (recovery times from 5 to 100 days).

It is less clear how to relate $P_I$ to observable in the real world. In the model, $P_I$ represents the probability that one node becomes infected if it has been in contact with another infected node in a chosen time interval. This number depends on many variables in the real world, such as how strong the contact is, which kind of interaction the nodes have, the viral load, etc. Taking all these possibilities into account would result in a more realistic model. At this stage, we opt for a simpler model with a single value for $P_I$ and for using the same range for both $P_I$ and $P_R$.

A posteriori, we discovered that $P_I$ is inversely proportional to the key Gompertz function time-constant $U$, and changing $P_I$ can be also be interpreted as a rescale of the time.

**Assessing whether a single trajectory follows the Gompertz growth function.**

We consider the function $J(t) = \ln\left[\ln\left[\frac{X(t)}{X(t-1)}\right]\right]$. Elementary algebra shows that $J(t)$ is a linear function if and only if $X(t)$ is the Gompertz growth function. Assessing the linearity of $J(t)$ is then sufficient to determine if the simulated infections are growing as the Gompertz function. By visual inspection of the $J(t)$ function, we can observe that in the vast majority of cases, the infection spread like Gompertz function in scale-free networks, but not with other networks. The results were further confirmed by computing the correlation coefficient r, which is always close to 1. The only cases in which the growth is not Gompertz are cases in which the virus disappears before being able to infect the network (<0.5% of the network is infected). The same is not true for other networks (**Figure 5** and **Supplementary Figure 1**).

**Proportionality between the newly daily infected and the "surface" of the epidemics.**

We define *C(t)* as the number of nodes that are connected to at least at one node in *I(t)* (i.e. nodes that have been infected but did not recover yet). Each node in *C(t)* can be connected t to multiple nodes in *I(t)*. We call $r_i$ the number of connections the i-th node belonging to *C(t)* that are received from all the nodes in *I(t)*.

From $r_i$, we can derive an approximate formula to derive the probability that the i-th node is infected. The probability that the *i*-th node will not be infected at a certain step *t+1* is indeed:

$$P_0 = (1 - P_I)^{r_i}$$

So that the probability of being infected, for the same node, is:

$$P_{INFECTION} = 1 - P_0 = 1 - (1 - P_I)^{r_i} \approx r_i P_I$$

Where the last approximate equality holds if $P_I \ll 1$.

The total nodes that are infected at time *t+1* are then:

$$\Delta(t+1) = P_I \cdot \sum_{i \in C(t)} r_i = P_I \cdot \Sigma(t)$$

Where we defined the quantity *Σ(t)* as the total number of contacts that each infectious node makes with non-infected nodes (this is the "surface" of the epidemic).

## FIGURES

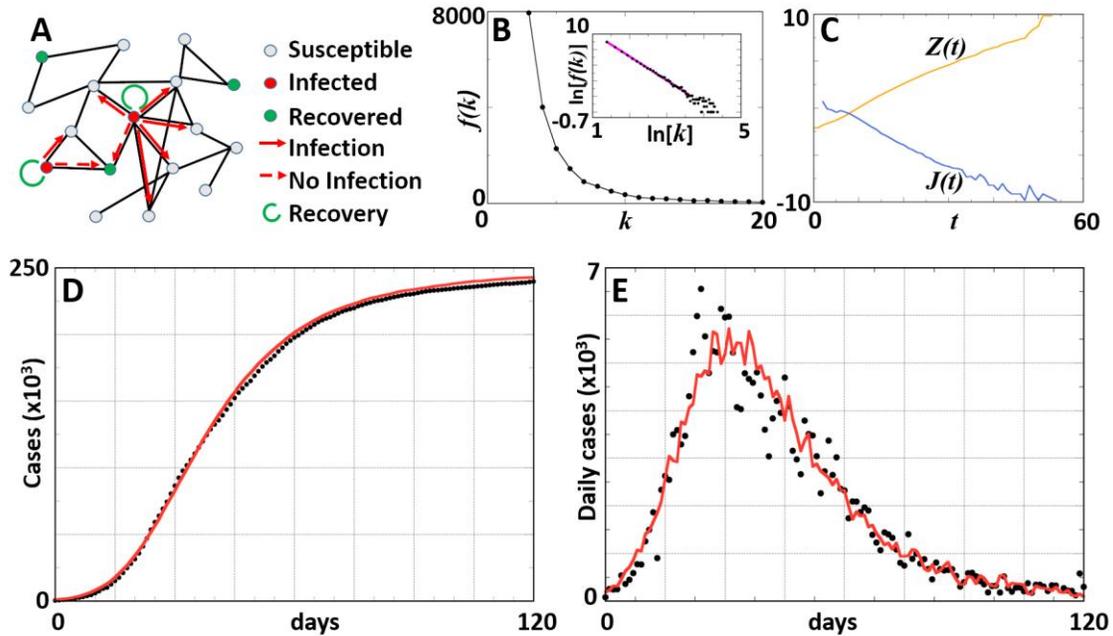

**Fig. 1. Spread of the virus modeled on a network. (A)** Our model of virus spread: at each time step, infected nodes (in red) can transmit the infection to a susceptible node (grey) with a probability $P_I$, only if there is a connection between the two nodes. After the infection phase, infected nodes can recover with a probability $P_R$. Recovered nodes cannot be re-infected. Panel **(B)** shows the distribution of the ranks of the nodes in the scale-free network used in the simulations. The inset shows the same plot on an ln-ln scale. As expected from theory (11), there is a straight-line fit with exponent (or slope) γ=-3.0. Panel **(C)** Shows the functions $J(t)=\ln[\ln[X(t)/X(t-1)]]$ and $Z(t)=\ln[\ln[N/X(t)]]$ (where $X(t)$ is count at time $t$; $N$ is plateau or maximum count level) during a simulated infection with $P_I$ =0.1 and $P_R$ =0.1. The fact that functions $J(t)$ and $Z(t)$ are linear functions of time indicates that viral spread follows Gompertz growth. The slopes of the $J(t)$ and $Z(t)$ lines are equal to $-1/U$ and $1/U$, where $U$= 4.7 & 4.6, days for $J(t)$ and $Z(t)$, respectively. Gompertz growth is observed only when the network is scale-free (**Supplementary Figure 1**). Panels **(D)** & **(E)** show a comparison between the results produced by a simulation ($P_I$ =0.02 and $P_R$ =0.02, red lines) and the actual confirmed cases observed in Italy (total and daily respectively, black dots; for the first outbreak). Since the network used in the simulations is relatively small (20,000 nodes), results need to be rescaled by a factor of 13.4 for the comparison to real data.

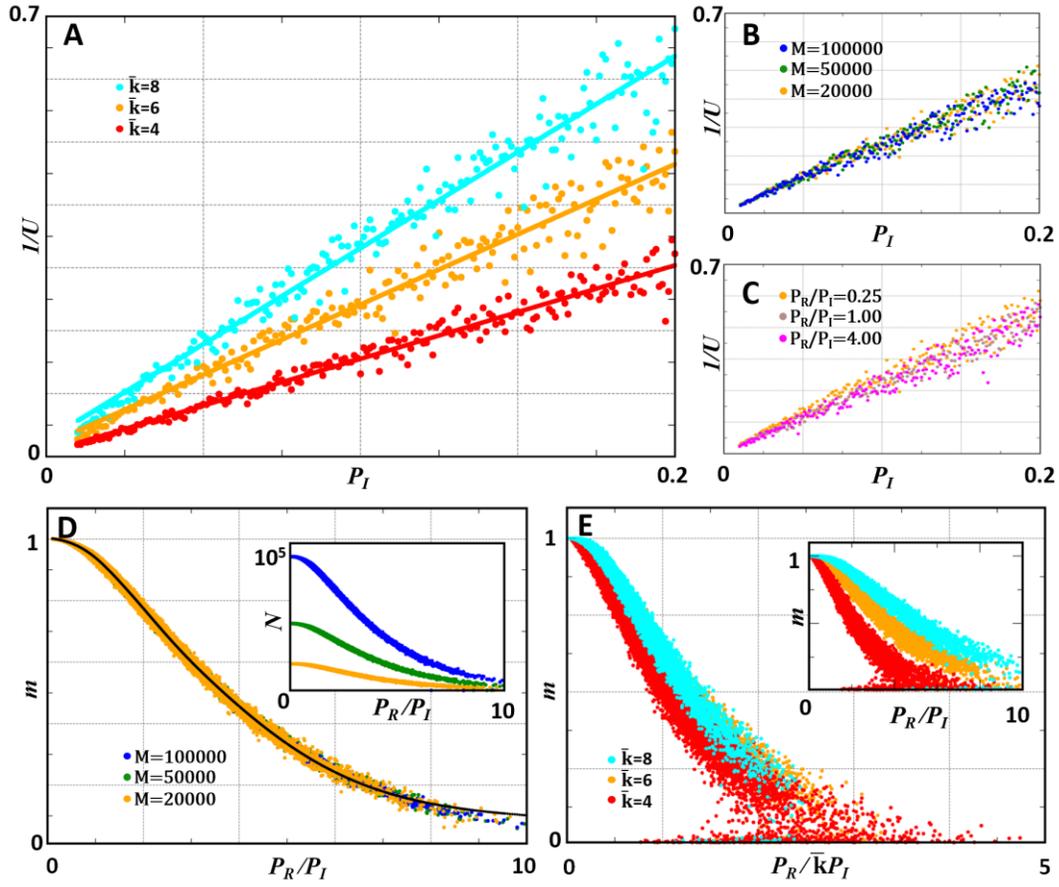

**Figure 2. The speed of spread and the fraction of the network that is infected at the plateau depend on the microscopic properties of the virus infectivity and on average network connectivity.** Panel (**A**): the parameter $1/U$ is a linear function of $P_I$ with slope proportional to the average number of contacts of the network, $\bar{k}$. The three sets of data correspond to three networks with $\bar{k}=4$ (red dots), $\bar{k}=6$ (orange dots), and $\bar{k}=8$ (cyan dots) and M=20,000 nodes. Straight lines are best linear fit and have slopes of 1.49, 2.23, and 3.05, respectively, making the value for the constant $\alpha$ in Eq. (2) equal to 0.375. Panels (**B**) and (**C**) show that $1/U$ does not depend on the size of the network, $M$, or the recovery probability, $P_R/P_I$. Panel (**D**) shows the fraction of the network that is infected $m=N/M$ (where $M$ is the total number of nodes in the network), depends on $P_R/P_I$ for a fixed value of $\bar{k}(=6)$. Note that $m$ does not depend on the size of the network. The black line is the fit of $m$ as a function of $P_R/P_I$ given in Eq. 3. The first four terms achieve a good fit, with the following values for the constants: $a_1=0.20$, $a_2=0.68$, $a_3=0.10$, $a_4=0.02$, $c_1=0.07$, $c_2=0.05$, $c_3=0.11$, $c_4=0.42$. The inset shows the plateau values $N$ as a function of $P_R/P_I$ for different values of $M$. Panel (**E**) shows how $m$ changes as a function of $\bar{k}$. The main graph shows that if $P_R/P_I$ is rescaled by $\bar{k}$, we obtain the same dependence on $m$, whereas the inset shows the functions before rescaling.

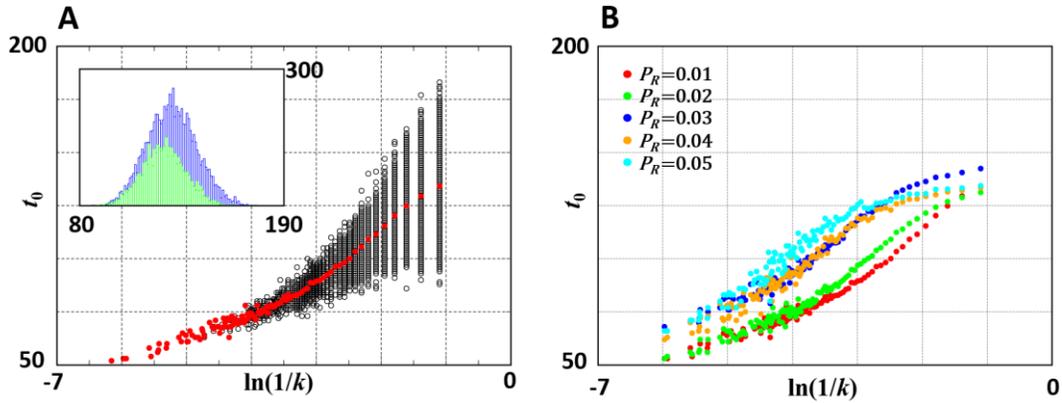

**Figure 3. Larger nodes are infected earlier in the epidemic.** Panel (**A**) shows $t_0$, the time at which the node is infected as a function of the degree of the node (more precisely $-\ln[k]$). Each black dot represents a different specific node. The time of infection is averaged over 100 replica simulations. The $t_0$ value averaged over all the nodes of same degree $k$ is shown by the red line. The inset shows histograms of the average time of infection for nodes with $k=3$ (blue) and $k=4$ (green); both distributions are normal distributions. Panel **B** plots the average $t_0$, value for various $P_R$ values when $P_I$ is set to 0.01 (we average over all 100 replicas and over all the nodes with the same degree, $k$). The red dots correspond to the red line in panel A. Differences in $t_0$ are less evident with higher values of $P_I$ (see Supplementary Figure 3).

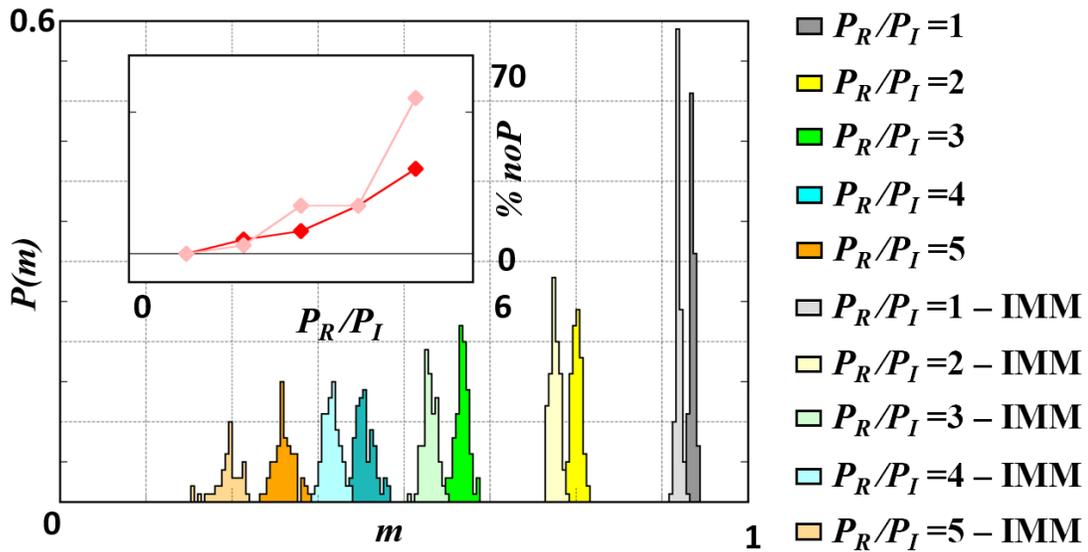

**Figure 4. Granting immunity to larger nodes reduces the fraction of the network that is infected.** Each histogram shows the plateau fraction $m=N/M$ for a set of 100 simulations; for different values of the ratio $P_R/P_I$ and for two different conditions: (*i*) every node can be infected, (*ii*) nodes with most neighbors (top 0.1% of all nodes) are immune from the beginning of the simulation. The inset shows the percentage of simulations in which the virus does not spread into the network, i.e., the final fraction of *m* is below 0.005. Immunization of highly connected nodes decreases the plateau value of *m* and increases the probability that the virus does not spread in the network.

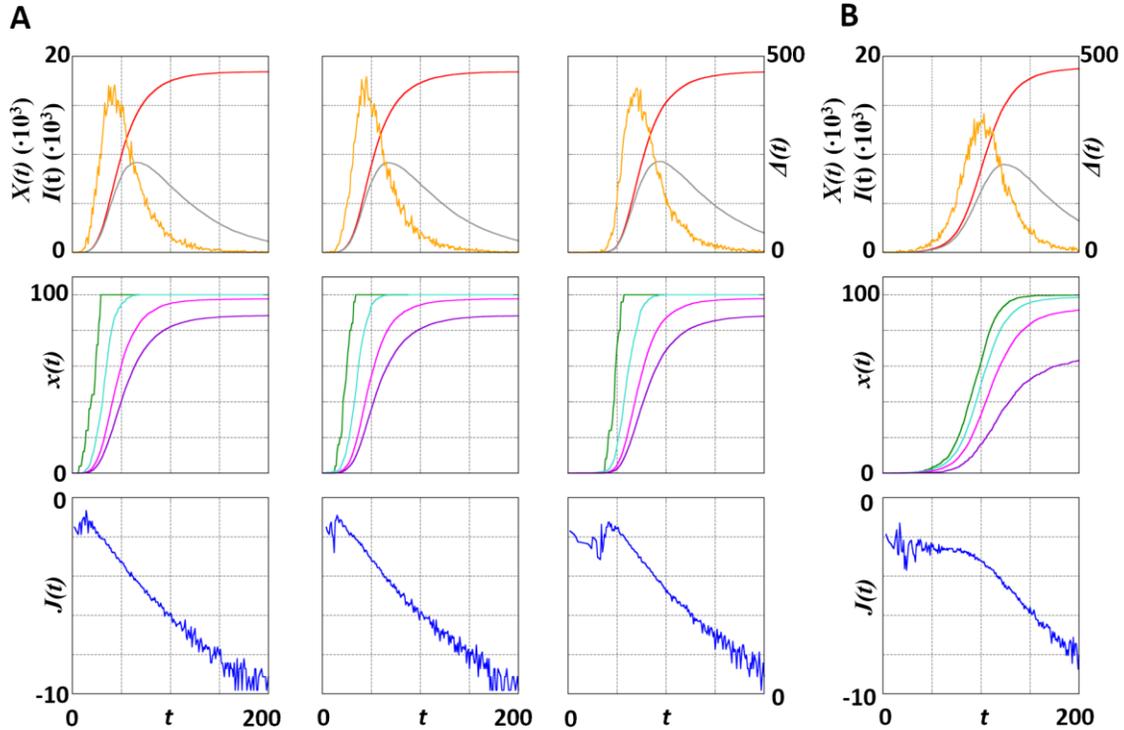

**Figure 5. In a scale-free network, rapid spread of the virus is caused by the infection of the larger nodes.** Panel (A) shows three replica simulations for the scale-free network topology, whereas panel (B) shows one typical simulation for the random network topology. In all the cases, we ran simulations with $P_I$ =0.02 and $P_R$ =0.02. The top row plots functions $X(t)$ (red line), $I(t)$ (gray line) and $\Delta(t)$ (orange line) against time, $t$. The middle row plots the percentage of infected nodes as a function of $t$ for four ranges of $k$ value (we called this function $x(t)$). For the scale-free network, the clusters are $k \geq 100$ (~0.1% of the total, dark green line), $20 \leq k < 100$ (2.7% of the total, cyan line), $5 \leq k < 20$ (37.2% of the total, magenta line) and $k < 5$ (60.0% of the total, purple line). For the random network, the clusters, which involve much smaller $k$ values, are $k \geq 10$ (8.5% of the total, dark green line), $6 \leq k < 10$ (47.1% of the total, cyan line), $3 \leq k < 6$ (38.2% of the total, magenta line) and $k < 3$ (6.26% of the total, purple line). Most remarkable is that in the scale-free network, the largest nodes are fully infected very early and before the next largest nodes. In the random network, the largest nodes are infected a little before smaller nodes but they take much longer to be fully infected. The bottom row plots the functions $J(t)$ (ln (exponential growth rate), blue line).

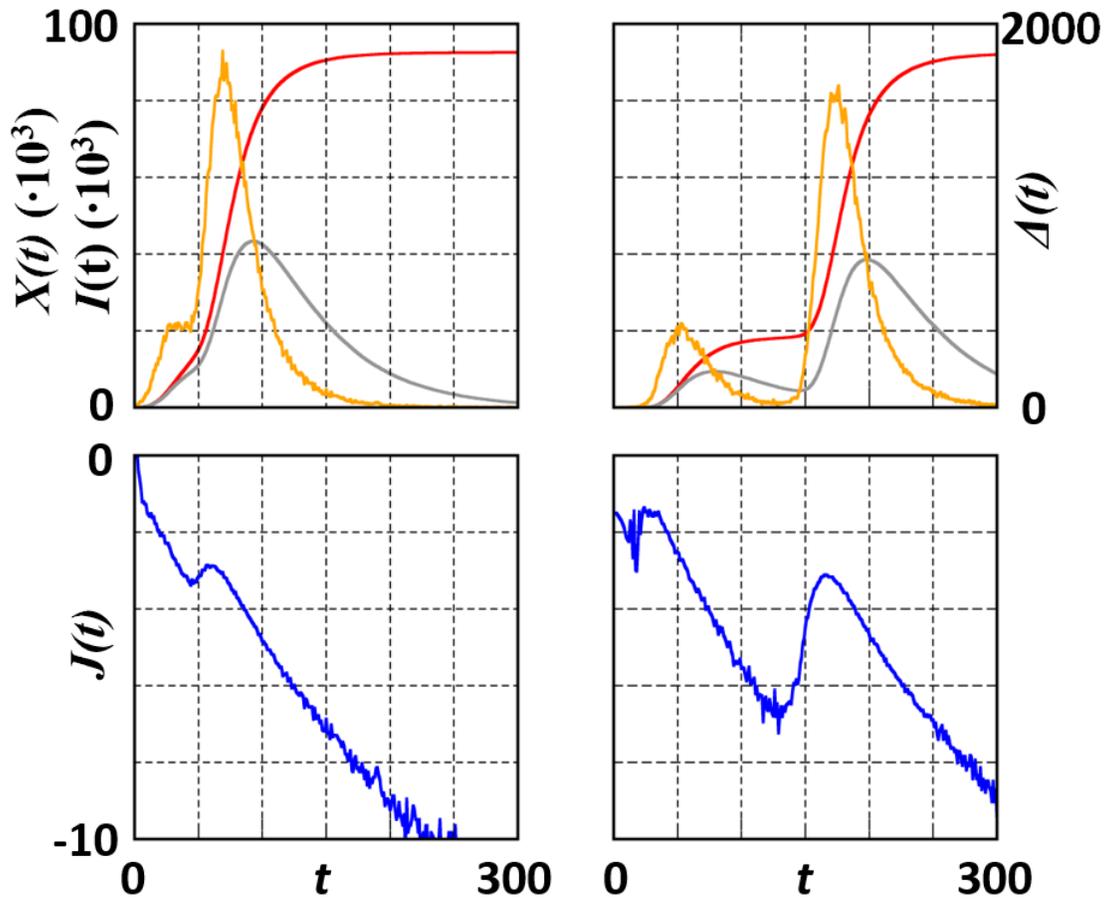

**Figure 6. Multiple outbreaks emerge in networks built from connected scale-free networks.**

We simulated the infection in a network that consist of two scale-free network ($M_1$=20000, $M_2$=100000) connected by a small number (10) of edges. In both cases we infected 4 random nodes of the smaller network at the beginning of the simulations. The infection and recovery rate are fixed at $P_I$ =0.02 and $P_R$ =0.02. The top row plots functions $X(t)$ (red line), $I(t)$ (gray line) and $\Delta(t)$ (orange line) against time, $t$; the bottom row plots the functions $J(t)$ (ln (exponential growth factor), blue line).

The two replicas show that the infection of the second network can happen at different times: in the first case we almost have superposition of the two peaks, while in the second case they are well separated in time. The function $J(t)$ resembles the one we observed in the real world data for multiple outbreaks.

**SUPPLEMENTARY FIGURE CAPTIONS**

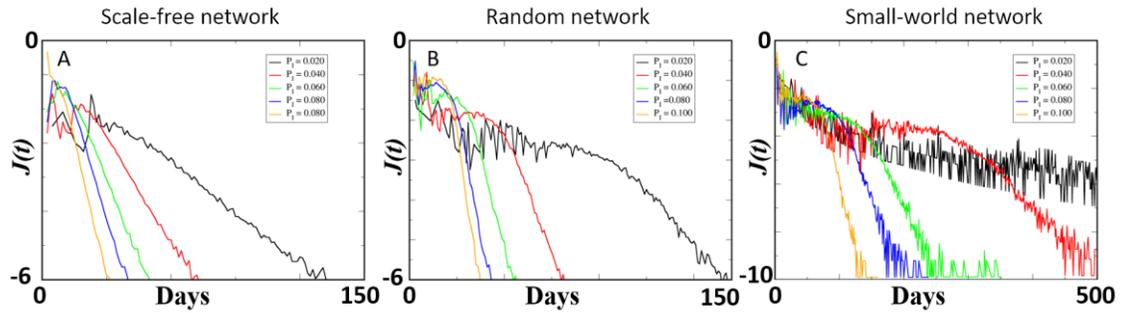

**Figure S1. Only in a scale-free network does the virus spreads according to the Gompertz law.** In each panel we show typical simulations with five different infectivity rates, $P_I$. The three panels compare virus spread in a network with different topologies: (A) scale-free network, (B) random network, and (C) small-world network, for different values of the infection rate $P_I$. Function $J(t)$ is linear only in the scale-free network, proving that the growth function is a Gompertz function

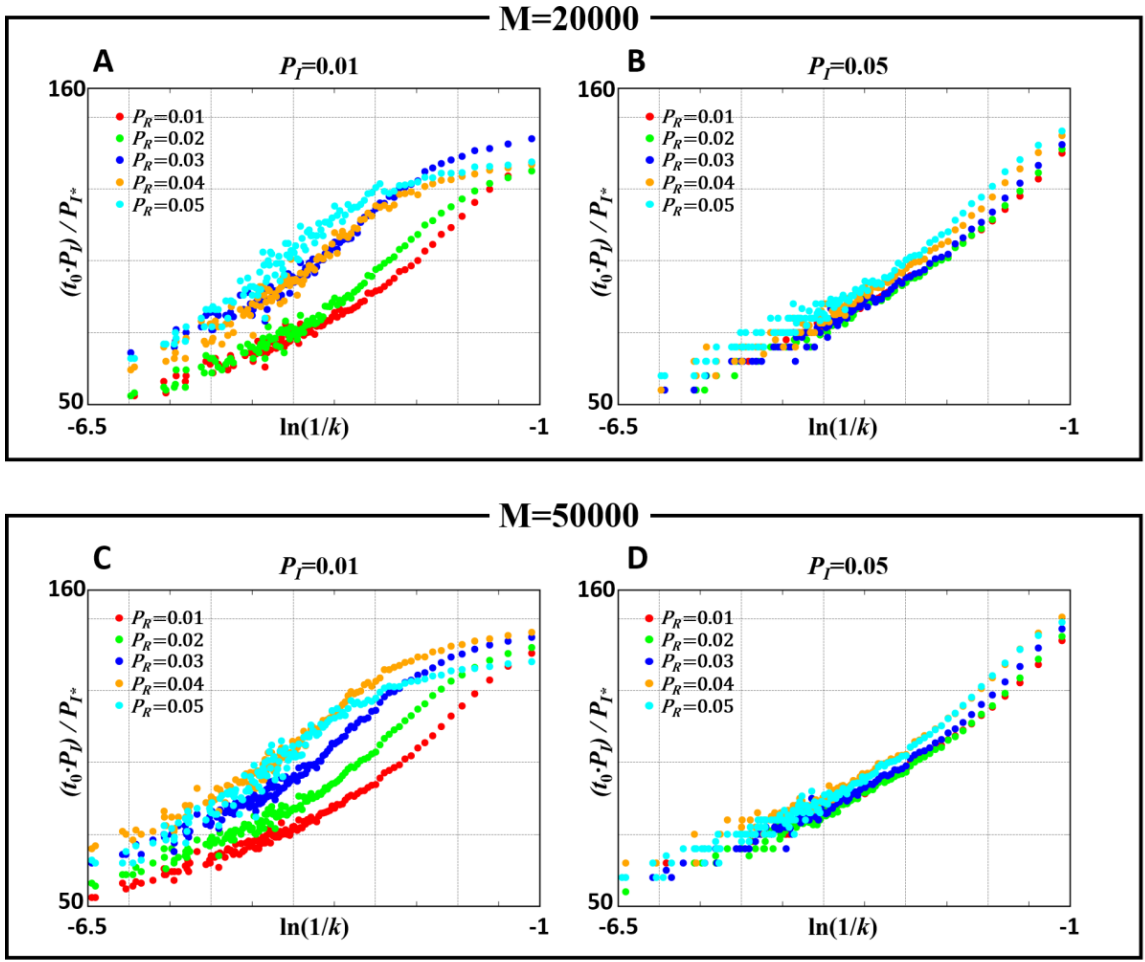

**Figure S2.** Time of infection for each node averaged over 100 replicas for all the nodes with the same degree $k$. We compare two different values of $M$ (20,000 and 50,000) and two different values of $P_I$ (0.01 and 0.05). Panel **A** is the same as panel **B** in **Figure 3**. The time on the Y-axis is normalized to the value of $P_I$, the reference value being $P_I^*$=0.01.

**ACKNOWLEDGEMENTS**

We thank Dr. Steven Strong for contacting us and sharing his idea of simulating virus spread on a network. His preliminary results convinced us to do the full-scale study that led to this paper.

Support was provided by US National Institutes of Health award R35GM122543 (M.L.) and National Natural Science Foundation of China Grant No. 31770776 (F.Z.). Michael Levitt is the Robert W. and Vivian K. Cahill Professor of Cancer Research.